\documentclass[10pt,twocolumn,letterpaper]{article}

 \usepackage[pagenumbers]{cvpr} 

\definecolor{cvprblue}{rgb}{0.21,0.49,0.74}
\usepackage[pagebackref,breaklinks,colorlinks,allcolors=cvprblue]{hyperref}

\def\paperID{*****} 
\def\confName{CVPR}
\def\confYear{2025}

\title{FEDLAD: Federated Evaluation  of Deep Leakage Attacks and Defenses}

\author{Isaac Baglin, Xiatian Zhu and Simon Hadfield\\
Centre for Vision, Speech and Signal Processing\\
University Of Surrey\\
Guildford GU2 7XH \\
{\tt\small \{ib00304,xiatian.zhu,s.hadfield\}@surrey.ac.uk}
}

\begin{document}
\maketitle
\begin{abstract}
Federated Learning is a privacy preserving decentralized machine learning paradigm designed to collaboratively train models across multiple clients by exchanging gradients to the server and keeping private data local. Nevertheless, recent research has revealed that the security of Federated Learning is compromised, as private ground truth data can be recovered through a gradient inversion technique known as Deep Leakage. While these attacks are crafted with a focus on applications in Federated Learning, they generally are not evaluated in realistic scenarios. This paper introduces the FEDLAD Framework (Federated Evaluation of Deep Leakage Attacks and Defenses), a comprehensive benchmark for evaluating Deep Leakage attacks and defenses within a realistic Federated context. By implementing a unified benchmark that encompasses multiple state-of-the-art Deep Leakage techniques and various defense strategies, our framework facilitates the evaluation and comparison of the efficacy of these methods across different datasets and training states. This work highlights a crucial trade-off between privacy and model accuracy in Federated Learning and aims to advance the understanding of security challenges in decentralized machine learning systems, stimulate future research, and enhance reproducibility in evaluating Deep Leakage attacks and defenses.
\end{abstract}    
\section{Introduction}
\label{sec:intro}
Marked by the proliferation of IoT devices, concerns regarding privacy, and information leakage have spurred the enactment of data protection legislation worldwide, highlighted by the GDPR in the European Union, and the Data Protection Act in the UK \cite{gdpr,UKData}. Traditional centralized machine learning techniques face challenges in safeguarding privacy during data collection and sharing, prompting the rise of Federated Learning (FL) as a privacy-preserving distributed learning paradigm \cite{mcmahan2017communication}. FL enables collaborative model training among participants without the exchange of local private data, a pivotal shift away from centralized methods. It operates by distributing the current model to each client, where local data is utilized for training. Gradient or model updates are subsequently shared with a central server for aggregation, as exemplified in algorithms like FedSGD and FedAvg respectively. This distributed approach not only addresses privacy concerns but also enhances scalability and security by minimizing data transmission, storage, and management overhead. FL finds application across diverse domains, from healthcare to autonomous vehicles, enabling the development of robust machine learning models while seemingly respecting data privacy and security regulations\cite{rieke2020future,pokhrel2020decentralized}.

Unfortunately, FL instils a misleading sense of security as in recent years a gradient inversion attack called Deep Leakage has emerged \cite{zhu2019deep}. This attack aims to uncover private training data by analyzing the shared gradients transmitted to the server by the client. The attacker, often a curious server or a third party intercepting the communications, employs a randomly initialized dummy image and the shared global model to compute dummy gradients. The goal is then to minimize the distance between these dummy gradients and the intercepted shared gradients, thus optimizing the dummy image to closely resemble the ground truth. Although Deep Leakage was initially proposed within the context of Federated Learning, early iterations of the attack were primarily limited to untrained networks \cite{zhu2019deep,zhao2020idlg}. Subsequent endeavours have aimed for more realistic scenarios by incorporating gradients from trained networks; however, these efforts typically assume a single steady-state model, failing to sufficiently consider how the recovery changes during training \cite{geiping2020inverting,yin2021see}. In a realistic scenario with a sustained attack being undertaking throughout the training process the attacker can theoretically achieve more accurate recoveries\cite{geng2110towards,geng2023improved}.
We call these approaches which leverage information from multiple timestamps Multi-Observation attacks.

Previous Deep Leakage techniques have been assessed using differing metrics, across dissimilar datasets, and employing divergent attack protocols \cite{zhu2019deep,geiping2020inverting,geng2110towards}. 
Prior codebases have utilized pretrained models in a stable state and lack implementation within a Federated Context or have minimal evaluation protocols in place\cite{Breaching,repotakahashi2023aijack,takahashi2023aijack}. In this paper, we present the FEDLAD Framework (Federated Evaluation of Deep Leakage Attacks and Defenses), which offers a diverse set of evaluation metrics across various datasets, incorporating implementations of numerous state-of-the-art attacks and defenses, and is designed to be extensible for future research. We hope this will help stimulate improved growth in the field, and better allow reproducibility in the future.

In FEDLAD, we consider a server that is honest-but-curious. This server conducts training as intended but, in its curiosity, collects gradients at different timestamps to construct its own dataset using the recovered ground truths. To be clear, the attack does not adversely affect the performance of the final FL model, but rather leaks the private data held by the participants. Our threat model in our benchmark assumes the attacker only has access to information known to the FL server and does not allow the attacker to modify the model structure as this would alert the victim of the attack. Auxiliary information is permissible when derived from commonly shared information, such as batch normalization statistics from global models \cite{yin2020dreaming}. Finally, the attacker cannot know the quality of the recovery as they don't have access to the ground truth. The code for our benchmark is available in the supplementary material.

Our main contributions in this paper are as follows:
\begin{itemize}
\item We present the FEDLAD Framework designed to assess single-observation and Multi-Observation attacks within the context of Federated Learning.

\item 
We unify various state-of-the-art attacks into a single joint formulation, and demonstrate the effectiveness of each approach across several widely recognised metrics for a variety of batch sizes, attack paradigms and defences.

\item We integrate FEDLAD with the popular Flower library for federated learning, and extend Flower to operate in FedSGD mode. We release all relevant code, ensuring that future works can also compare fairly and benefit from the new insights highlighted by the benchmark.
\end{itemize}
\section{Related Work}
\label{sec:formatting}

Gradient inversion attacks were first formulated in 2019 by Zhu et al. with an attack called Deep Leakage from Gradients (DLG) \cite{zhu2019deep}. This attack tries to minimize the Euclidean distance between the dummy gradients generated using a randomly initialized image and the real gradients taken by the attacker. L-BFGS is used to optimize the dummy image based on the Euclidean distance calculated in each iteration \cite{journals/mp/LiuN89}. It achieved successful image reconstruction for the MNIST, CIFAR, SVHN, and LFW datasets however they were only able to realize their attack using gradients sourced from randomly initialized network weights\cite{cifar,deng2012mnist,LFWTech,netzer2011reading}. Zhao et al. proposed `Improved Deep Leakage From Gradients (IDLG)' which builds upon the initial attack by extracting the ground truth label from the gradients of the last fully connected layer \cite{zhao2020idlg}. Since this attack doesn't require optimizing the label jointly, it exhibited a faster convergence speed than DLG. IDLG was limited to a batch size of 1 and, akin to DLG, exclusively operated with randomly initialized weights.These limitations make it unrealistic to apply the attack in a real FL setting.

Geiping et al. put forward Inverting Gradients which instead used the cosine similarity as their cost function and the Adam optimizer to update the dummy image \cite{geiping2020inverting}. In contrast to the prior attacks, Inverting Gradients used total variation to regularize the recovery. With these advancements, they managed to effectively execute attacks on trained networks, which usually exhibit gradients of varying magnitudes and tend to be harder to recover. A localization effect was identified where the object within an image would shift around when using trained networks. DeepInversion has more advanced regularization on by incorporating $\ell_{2}$ norm alongside a novel BN regularization term that leverages feature maps' mean and variance \cite{yin2020dreaming}. Yin et al. integrated these regularization components into their GradInversion attack, introducing a unique group consistency regularization term \cite{yin2021see}. This term optimizes multiple random seeds simultaneously to generate a consensus image, thereby allowing the attacker to penalize outliers. Gradinverion was found to be very effective for larger batch sizes and more advanced datasets such as Imagenet \cite{deng2009imagenet}. Despite these improvements, Gradinversion fails if there are multiple images of the same class within a batch \cite{dimitrov2022data}. 

 Geng et al. introduced the first Multi-Observation attack through their proposal of Group Consistency from Multiple Updates as well as a zero-shot batch label recovery that isn't limited by the label distribution \cite{geng2110towards}. Their attack involves selecting weight and gradient pairs taken from various timestamps, computing the gradient distances with each pair and using the sum to update the dummy image. The attack was said to be more successful when using an increasing number of pairs. It would however be computationally unrealistic over prolonged training scenarios because computation scales with the number of weight and gradient pairs. Upon recovery, images frequently undergo shuffling within a batch; hence, PSNR is employed to align them with ground truths before evaluation, enabling them to quantify the recovery quality \cite{geng2023improved}. In our work, we utilize  a similar approach with the LPIPS metric because it is more robust to the localization challenges observed in previous studies\cite{DBLP:journals/corr/abs-1801-03924lpips,geiping2020inverting}.
 
Yang et al. classify all the above Deep Leakage attacks as "optimisaiton based"\cite{yang2023gradient}. In contrast, they define "analytics-based" methods, which are those that obtain the ground truth data by solving a linear system of equations. Fowl et al. propose an analytics-based attack which side-connects a large fully connected network to the original global model to obtain a copy of the input data directly \cite{fowl2021robbing}. As this method extends the global model, the victim would be aware of the attack, and it is arguably impractical.

As in other areas of machine learning, the demonstration of novel AI attacks and defences are inextricably linked. The initial Deep Leakage paper discussed several simple methods to protect a system
from Deep Leakage; noisy gradients, gradient pruning, increasing the batch size and using higher
resolution images \cite{zhu2019deep}. Such methods are often referred to as Differential Privacy (DP). PFMLP was introduced by Fang et al. which encrypts gradients using homomorphic encryption before transmission \cite{fang2021privacy}. While they assert that accuracy deviation is limited to approximately 1\%, they acknowledge that their encryption process substantially escalates computational overhead. To address this challenge, they implement an enhanced Paillier algorithm in an attempt to mitigate the impact.

In 2021, Zheng et al. found that using additional dropout layers before the classifier was very effective at preventing Deep Leakage however much like previous methods it came at the cost of model performance\cite{zheng2021dropout}. The following year, Scheliga et al. proposed the Dropout Inversion Attack(DIA) \cite{scheliga2023dropout}. Dropout masks are initialized randomly from a Bernoulli distribution and are jointly optimized with the dummy data to approximate the client's model realization. They found that dropout gives a misleading sense of security, as their attack sometimes surpassed scenarios where the model had no dropout applied. In a separate paper, Scheliga et al. introduced PRECODE as a model extension to safeguard gradients against Deep Leakage by concealing the latent feature space through variational modelling \cite{scheliga2022precode}.
More recently, in 2023, Gao et al. used carefully selected transformations on sensitive images to make it more challenging for an adversary to recover them from the gradients \cite{gao2023automatic}. Their system automatically selects the transformation policies which minimise the recovery and maximise the model accuracy. Finally, DCS optimizes a dummy image to generate gradients that stimulate the same response as the client's gradients, while also maximizing the difference between the dummy image and the ground truth \cite{wu2023concealingsensitivesamplesgradient}. The evaluation of both attacks and defenses lack consistency, as they are analyzed across various training states, utilizing diverse model architectures, and employing different metrics. Our benchmark facilitates comprehensive evaluations of state-of-the-art Deep Leakage attacks, and it can be expanded to accommodate future developments in the field.
\section{Method}

Our paper indroduces the FEDLAD Framework designed to test the efficacy of a broad range of Deep Leakage attacks in a consistent manner. Our framework introduces a range of new insights into the success of State-Of-The-Art attacks when assessing their ability to recover ground truth information. The evaluation introduces a wide range of features including novel metrics which couldn't previously be considered due to the limitations of prior work. This enables robust assessment of potential privacy breaches in Federated Learning systems and also allows future work to evaluate defensive strategies. Moreover, within this framework we propose an attack that leverages Multi-Observation information in a scalable manner. We hope our unified and extensible codebase will make it simple for future research to adopt this more realistic attack framework. Our codebase uses elements of the AIJack library (Apache License 2.0) as a skeleton for the attacks and the TorchMetrics library (MIT License) for implementing prior metrics and Flower for FL functionality \cite{takahashi2023aijack,Detlefsen2022,beutel2020flower}. 
\subsection{FEDLAD Framework}

The Evaluation protocol within FEDLAD encompasses a range of attack metrics and scenarios, all outlined in Table \ref{sample-table3}. The selection of Evaluation Features is very inconsistent in prior work and doesn't offer an easy comparison between attacks (Table \ref{sample-table}). Our framework facilitates comparisons of prior attacks by re-implementing them all in a consistent codebase, while also introducing three novel considerations: Multi-Observation analysis, computational cost, and  Recovery Consistency Index (RCI) (Table \ref{sample-table2}). Multi-Observation analysis considers the repetition of attacks at various timestamps during training rather than the traditional approach in the field of only attacking once during FL. Certain attacks leverage multiple gradients, weights, or recoveries from these timestamps to enhance performance. Otherwise, this feature allows us to analyse the recovery quality throughout training. Secondly, previous studies often overlook the computational cost when presenting their attack strategies. This becomes increasingly important with certain Multi-Observation attacks that scale exponentially during the FL process. Hence, FEDLAD elucidates the trade-off between computation and recovery. The quality of recovery relies on the magnitude of the gradients, thus the recovery quality is directly influenced by the training state of the global model. As such, we propose a new metric, Recovery Consistency Index (RCI) which quantifies the recovery quality across the entire training process rather than only the trained and untrained models as in previous work. To determine RCI, we compute the area-under-the-curve of the recovery performance throughout the entire training process as:
\begin{equation}
     RCI = \frac{\Delta \mathcal{I}}{\mathcal{I}_{N}}\left( \frac{\mathcal{R}(\mathcal{I}_{N}) +  \mathcal{R}(\mathcal{I}_{0})}{2} +\sum\limits_{t=1}^{N-1} \mathcal{R}({\mathcal{I}}_{t})\right),
\end{equation}
where $R(\mathcal{I})$ is the recovery quality at training iteration $\mathcal{I}$ for $N$ timestamps taken every attack rate $\Delta \mathcal{I}$ . The recovery metric $R(\mathcal{I})$ employed to compute the RCI is Learned Perceptual Image Patch Similarity (LPIPS), where a lower value indicates better recovery, thus a lower RCI reflects greater consistency in recovery throughout the entire training process. We can thus interpret RCI as the expectation of the recovery quality for an attack performed at an arbitrary point during FL training. 

\begin{figure*}
    \centering
    \includegraphics[width=0.8\linewidth]{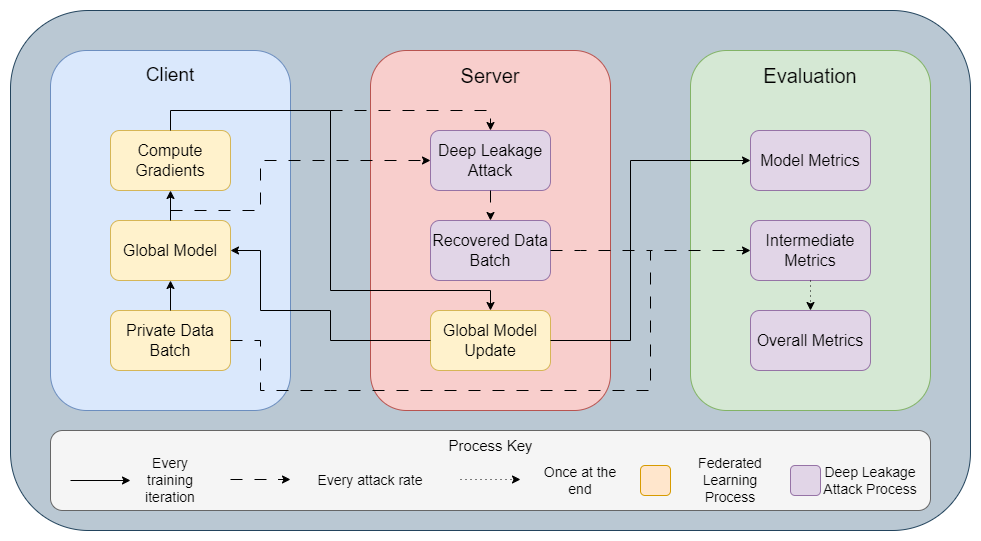}
    \caption{Overview of the interconnected training, attack, and evaluation processes that mutually influence each other during the training phase.}
    \label{fig:enter-label}
\end{figure*}

Within FEDLAD we define 3 types of metrics; Model, Intermediate and Overall (Figure \ref{fig:enter-label}). Model metrics are calculated at the end of each training iteration to quantify the model loss and accuracy. This is significant, as many defenses against Deep Leakage involve a trade-off between privacy and performance.The exact form of these metrics depends on the specific model being trained. Intermediate metrics are those which are computed after each attack to determine the recovery quality of that attack. Within this category, Mean Squared Error (MSE), quantifies the average squared difference between the recovered and ground truth images, emphasizing outliers. Learned Perceptual Image Patch Similarity (LPIPS), captures perceptual similarity by considering local image patches, aligning more closely with human visual perception. Structural Similarity Index Measure (SSIM), evaluates the similarity between two images by analyzing luminance, contrast, and structure, providing a holistic view of image quality. Peak Signal-to-Noise Ratio (PSNR), measures the ratio of the maximum possible signal power to the power of corrupting noise, offering insights into the fidelity of reconstructed images. The final category are overall metrics which are calculated at the end of the training process to assess the overall recovery quality throughout the entire training duration,  specifically the RCI metric.

One of the primary obstacles in recovering ground truth data is the tendency for images of the same class to be rearranged between iterations, with the original sequence dictated by the initial label recovery process \cite{geng2023improved}.This occurs because the gradients resulting from a model evaluation are agnostic to the ordering of the batch. This complicates the computation of metrics because naive direct comparisons between batches will lead to poor scores even for perfect but shuffled outputs. To tackle this issue, in our framework the Hungarian algorithm is utilized to match the recovered and ground truth images, with LPIPS serving as the metric for the cost matrix. Prior studies utilize PSNR as their metric for the cost matrix, which is inadequate since it doesn't consider localization changes between the ground truth and recovery \cite{geng2110towards,geiping2020inverting}.

\begin{table}
  \caption{Evaluation Features of Deep Leakage attacks.}
  \label{sample-table3}
  \centering
   \begin{adjustbox}{max width=0.46\textwidth}
  \begin{tabular}{cc}
    \toprule

   Evaluation Features     & Description     \\
    \midrule
    Trained Model & Can gradients from trained models be targeted for attack? \\

    Computation  & Does the evaluation include the computational cost of the attack?\\
    MSE     & Is the MSE of the ground truth relative to the recovery considered?\\
    LPIPS     & 
Does it evaluate the perceptual similarity (LPIPS)? \\
    PSNR     & Is the PSNR between the ground truth and the recovery factored in?\\
    SSIM    & Does it evaluate the structural similarity (SSIM)?\\
    RCI     & Is the consistency of the recovery throughout training considered? \\
    Multi-Observation Analysis    & Can the attack be run at various points throughout training? \\
    Federated Context     & Is it within a Federated paradigm rather than a steady state? \\
    \bottomrule
  \end{tabular}
  \end{adjustbox}
\end{table}
\begin{table}
  \caption{The previous state-of-the-field. Evaluation Features published in the original implementations of the State-Of-The-Art Deep Leakage attacks. }
  \label{sample-table}
  
  \centering
  \begin{adjustbox}{max width=0.5\textwidth}
  \begin{tabular}{ccccc}
    \toprule
    \multicolumn{5}{c}{Attack Name}                   \\
    \cmidrule(r){2-5}
    Evaluation Features    & DLG\cite{zhu2019deep}    & Inverting Gradients\cite{geiping2020inverting} & GradInversion\cite{yin2021see}   & Multiple Updates\cite{geng2110towards}   \\
    \midrule
    Trained Model &  & \checkmark & \checkmark & \checkmark  \\
    Computation   & & & &  \\
    MSE     &\checkmark &\checkmark &  & \checkmark \\
    LPIPS     & &  & \checkmark & \checkmark  \\
    PSNR     & &\checkmark & \checkmark & \checkmark \\
    SSIM     & & &  & \checkmark \\
    RCI     & & &  &  \\
    Multi-Observation Analysis     & & & &  \\
    Federated Context     & & & & \\
    \bottomrule
  \end{tabular}
  \end{adjustbox}
\end{table}

\begin{table}
  \caption{The new state of the field: Evaluation Features considered by FEDLAD to analyse the State-Of-The-Art Deep Leakage attacks.}
  
  \label{sample-table2}
  \centering
  \begin{adjustbox}{max width=0.5\textwidth}
  \begin{tabular}{ccccc}
    \toprule
    \multicolumn{5}{c}{Attack Name}                   \\
    \cmidrule(r){2-5}
    Evaluation Features     & DLG\cite{zhu2019deep}    & Inverting Gradients\cite{geiping2020inverting} & GradInversion\cite{yin2021see}   & Multiple Updates\cite{geng2110towards}    \\
    \midrule
    Trained Model & \checkmark & \checkmark & \checkmark &  \checkmark \\

    Computation  & \checkmark & \checkmark & \checkmark &  \checkmark \\
    MSE     & \checkmark & \checkmark & \checkmark &  \checkmark \\
    LPIPS     & \checkmark & \checkmark & \checkmark &  \checkmark \\
    PSNR     & \checkmark & \checkmark & \checkmark &  \checkmark \\
    SSIM    & \checkmark & \checkmark & \checkmark &  \checkmark \\
    RCI    & \checkmark & \checkmark & \checkmark &  \checkmark \\
    Multi-Observation Analysis    & \checkmark & \checkmark & \checkmark &  \checkmark \\
    Federated Context     & \checkmark & \checkmark & \checkmark &  \checkmark \\
    \bottomrule
  \end{tabular}
  \end{adjustbox}
\end{table}

The following subsections will specify our unified gradient leakage formulation, within which the evaluated prior works have been reimplemented.

\subsection{Label Recovery}
 To recover the ground truth labels, the number of occurrences of label $n$ can first be calculated as:

\begin{equation}
\sum\limits_{b}y^{*}_{b,n} \approx \sum\limits_{b}p_{b,n} - \frac{\mathcal{B}\Delta W_{n}^{FC}}{\bar{O}_{b}} 
\end{equation}

where $p_{b}$ is the post-softmax probability of batch sample $b$, $\bar{O}_{b}$ is the mean of the input of the final fully connected layer and $\Delta W_{n}^{FC}$ is the sum of the client gradients in the last fully connected layer of each classes layer output. We refer to \cite{geng2023improved} for the full derivation of this. A label vector $\hat{y}$ can be constructed using the class counts with the correct number of occurrences for each class.

\subsection{Data Recovery}
Next, we attempt to recover the private client data from the shared gradients of the global model's loss function w.r.t. the server weights $W$. To do this, we use a randomly initialized synthetic batch $\hat{x}$ and the predicted label vector $\hat{y}$ to produce dummy gradients. The distance between the real FL client gradients and the dummy gradients is computed as: 

\begin{equation}\label{eq:loss_def}
    \resizebox{0.4\textwidth}{!}{$\mathcal{L}_{grad}(\hat{x},\hat{y};W,\Delta W) = \delta(\Delta W \mathcal{L}(\hat{x},\hat{y}),\Delta W^{*} \mathcal{L}(x^{*},y^{*})),$}
\end{equation}

where $\Delta W^{*}$ represents the gradient of the loss function w.r.t the client weights $W^{*}$ and $\delta$ is the distance function employed . $\Delta W^{*}$ is shared by the client node, after computation on its own private image batch $x^*$ and label batch $y^*$ (* indicates private). We assume the attack takes place before the global model is updated during aggregation, hence $W^{*}$ = $W$. The minimization of the distance between the real and dummy gradients means that our reconstruction task can now be solved through the optimization:

\begin{equation} \label{eq:optim}
     \resizebox{0.4\textwidth}{!}{$\hat{x}^*,\hat{y}^*  = arg\underset{\hat{x},\hat{y}}{min}\left(\mathcal{L}_{grad}(\hat{x},\hat{y};W,\Delta W) + \mathcal{R}_{Tot}(\hat{x}) \right),$}
\end{equation}

where $\hat{x}^*$ and $\hat{y}^*$ are the recovered image and label batch respectively. A general regularization term $\mathcal{R}_{Tot}(\hat{x})$ is introduced to the optimization process. The nature of this regularization varies depending on which of the evaluated attacks is being employed, but in general form, it can be expressed as:

\begin{equation}
\resizebox{0.42\textwidth}{!}{$\mathcal{R}_{Tot}(\hat{x}) = \alpha_{TV}\mathcal{R}_{TV}(\hat{x}) + 
\alpha_{\ell_{2}}\mathcal{R}_{\ell_{2}}(\hat{x})+\alpha_{BN}\mathcal{R}_{BN}(\hat{x}) +  \alpha_{Group}\mathcal{R}_{Group}(\hat{x})).$}
\end{equation}

Total variation regularization $\mathcal{R}_{TV}(\hat{x})$ penalizes the changes in intensity between neighbouring pixels, promoting smoother images with fewer abrupt transitions \cite{RUDIN1992259}. The regularization term $\mathcal{R}_{\ell_{2}}(\hat{x})$ imposes regularization on $\hat{x}$ using the Euclidean norm, defined as the square root of the sum of squares of pixel intensities. This helps to reduce the occurrence of outlier pixel values. The BatchNorm regularization term $\mathcal{R}_{BN}(\hat{x})$ minimizes the difference between feature map statistics of $\hat{x}$ and $x^{*}$ and is formulated as:

\begin{equation}
\resizebox{0.42\textwidth}{!}{$\mathcal{R}_{BN} = \sum\limits_{l}^{L}\left(||\mu_{l}(\hat{x})-BN_{l}(Mean) ||_{2} + ||\sigma_{l}^{2}(\hat{x})-BN_{l}(Variance) ||_{2}\right),$}
\end{equation}

where $\mu_{l}(\hat{x})$ and $\sigma_{l}^{2}(\hat{x})$ are the mean and variance of the feature maps at convolutional layer $l$ of batch $\hat{x}$ with $BN_{l}(Mean)$ and $BN_{l}(Variance)$ representing the accumulated statistics computed during the training process. These can be recovered from the state dictionary of the BatchNorm layers \cite{yin2020dreaming}. The group consistency term $\mathcal{R}_{Group}(\hat{x})$ simultaneously optimizes multiple random seeds to produce a consensus image which enables the attacker to penalize deviations \cite{yin2021see}. Each regularization term possesses a scaling factor $\alpha$, which can be adjusted based on the chosen attack configuration (Table \ref{sample-table4}). All scaling factors are implemented to dynamically adjust based on both the batch size $\mathcal{B}$ and an image scaling factor $\mathcal{F}$. This factor $\mathcal{F}$ represents the image size ratio between the selected dataset and CIFAR10. For instance, if CIFAR10 is selected, $\mathcal{F}$ equals 1; whereas, if ImageNet is chosen, $\mathcal{F}$ equals 49.
 If the chosen attack mode utilizes gradients and weights sourced from different steps, the optimisation function, equation \ref{eq:optim} is reformulated as:

\begin{equation}
     \resizebox{0.42\textwidth}{!}{$\hat{x}^*,\hat{y}^*  =  arg\underset{\hat{x}\hat{y}}{min}\left(\sum\limits_{t=0}^{N}\left( \mathcal{L}_{grad}(\hat{x},\hat{y};W_{t},\Delta W_{t})\right)  + \mathcal{R}_{Tot}(\hat{x})\right)$} 
\end{equation}
where $N$ is the total number of steps $t$.
\begin{table}
  \caption{Reconstruction Features of the State-Of-The-Art Deep Leakage attacks.}
  \label{sample-table4}
  \centering
  \begin{adjustbox}{max width=0.5\textwidth}
  \begin{tabular}{ccccccc}
    \toprule
    \multicolumn{7}{c}{Reconstruction Feature}                   \\
    \cmidrule(r){2-7}
    Attack Name    & $ \alpha_{TV}$    & $\alpha_{\ell_{2}}$ & $\alpha_{BN}$   & $\alpha_{Group}$   & Optimizer & Distance $\delta$  \\
    \midrule
    DLG\cite{zhu2019deep}     &0 &0 &0 &0 & L-BFGS& $\ell_{2}$\\

    Inverting Gradients\cite{geiping2020inverting}  &0.08$\mathcal{F}$/$\mathcal{B}$  &0 &0 &0  &Adam & Cosine \\
    GradInversion\cite{yin2021see}    &0.08$\mathcal{F}$/$\mathcal{B}$ &0.0008$\mathcal{F}$/$\mathcal{B}$ & 0.0001$\mathcal{F}$/$\mathcal{B}$& 0.0001$\mathcal{F}$/$\mathcal{B}$ &L-BFGS  & $\ell_{2}$\\
    Multiple Updates\cite{geng2110towards}      &0.08/$\mathcal{B}$ &0 &0 &0 &L-BFGS &$\ell_{2}$ \\

    \bottomrule
  \end{tabular}
  \end{adjustbox}
\end{table}
\subsubsection{State-Of-The-Art Defense Implementations: Differential Privacy, PRECODE \cite{scheliga2022precode}, 
 DCS\cite{wu2023concealingsensitivesamplesgradient}}\label{sota}

The first defense we intend to evaluate is applying Gaussian noise to the gradients before they are exchanged with the server. The gradients are updated using:

\begin{equation}\label{eq:l}
    \Delta \hat{W}^{*} \mathcal{L}(x^{*},y^{*}) = \Delta W^{*} \mathcal{L}(x^{*},y^{*}) + \eta,
\end{equation}
where $\Delta \hat{W}^{*}$ is the noisy gradient and $\eta$ is a Gaussian distributed noise vector the same size as the client gradient $\Delta W^{*}$ with a standard deviation of $\alpha_{std}$. In our experiments we use 3 values for $\alpha_{std}$ 0.1, 0.01 and 0.001.

Secondly, we evaluate PRECODE \cite{scheliga2022precode} which is a model extension that adds a variational bottleneck before the final fully connected layer to conceal the latent feature space through
variational modeling. Given an input $i$, PRECODE first transforms it into a latent representation z using an encoder  $\mathcal{E}(\cdot)$ such that:

\begin{equation}
\mu,\sigma^2 = \mathcal{E}(i).
\end{equation}

Next, it uses the reparameterization trick: 
\begin{equation}\label{reparam}
z = \mu + \sigma\cdot\epsilon,
\end{equation}

where $\epsilon \sim \mathcal{N}(0, 1)$  is a random noise vector sampled from a standard normal distribution to calculate a latent vector $z$. A decoder $\mathcal{D}(\cdot)$ is then used to map the sampled latent representation $z$ back to the output space (the same size as the input) and is computed with: 

\begin{equation}
\hat{o} = \mathcal{D}(z), 
\end{equation}
 where $\hat{o} $ is the output of the variational bottleneck. The encoder and decoder in the PRECODE variational bottleneck are learnable linear layers. 
 
The final state-of-the-art defense implemented is DCS\cite{wu2023concealingsensitivesamplesgradient}, which is optimization based. They use equation \ref{eq:loss_def} to compute the difference between real and proxy gradients with their proxy gradients produced using a randomly initialized image $\tilde{x}$. They optimize $\tilde{x}$ so the gradients give similar responses such that:

\begin{equation} \label{eq:optim_def}
     \resizebox{0.42\textwidth}{!}{$\tilde{x}^*  = arg\underset{\tilde{x}}{min}\left(\mathcal{L}_{grad}(\tilde{x},\tilde{y};W,\Delta W) + \mathcal{L}_{dcs}(\tilde{x},x^*) \right),$}
\end{equation}
where $\mathcal{L}_{dcs}$ is a loss function used to ensure the $\tilde{x}$ and $x^{*}$ are dissimilar where $\mathcal{L}_{dcs}$ calculates the euclidean distance between $\tilde{x}$ and $x^{*}$. They use Adam as an optimizer with a learning rate of 0.1 and their chosen distance function is cosine similarity.
\section{Evaluation}

The FEDLAD framework was implemented using the Pytorch and Flower libraries and our experiments were run on an A100 GPU with 80GB of VRAM. Within our framework, we analyse 4 gradient inversion techniques with varying properties: Firstly the original DLG approach \cite{zhu2019deep} which was proposed for untrained models, secondly Inverting Gradients\cite{geiping2020inverting} which was able to operate on trained models, thirdly Gradinversion\cite{yin2021see} which was designed to handle larger batch sizes and finally Multiple Updates\cite{geng2023improved} which currently achieves state of the art accuracy but introduces additional assumptions and is computationally complex. The specific configurations of these attacks within our unified formulation are outlined in Table \ref{sample-table4}. Additionally, we consider 3 defence techniques: applying Gaussian Noise to the gradients on the client side, PRECODE\cite{scheliga2022precode} and DCS\cite{wu2023concealingsensitivesamplesgradient}. The evaluation of these attacks and defences takes place within a FedSGD classification training scenario, with the LeNet network utilizing CIFAR10 and ImageNet datasets. CIFAR10 is comprised of 50,000 images divided into 10 classes, each with an image size of 32x32 pixels. Conversely, ImageNet encompasses more than 1.2 million images spanning 1000 classes, cropped to a resolution of 224x224 pixels. In each experiment, the attack is executed at intervals of every 500 training iterations, spanning a total training duration of 10,000 iterations (learning rate of 0.01). Our framework has the option to use repeated batches for attacks or random. 
The Multiple updates approach requires repeated batches in order to function, thus to enable a fair comparisons, we use this throughout the experiments.
Each attack consists of 50 attack iterations. GradInversion simultaneously optimizes 6 random seeds for its group consistency term, while Multiple Updates uses 2 random seeds and chooses the seed with the lowest distance value to mitigate outliers. All experiments are repeated 5 times and we report the the arithmetic mean.

\begin{table}
  \caption{The effectiveness of various attack methods with a batch size of 8 for CIFAR-10 and ImageNet. Multiple Updates results omitted for ImageNet due to it exceeding computational limit of A100 GPU.}
  \label{sample-table5}
  \centering
  \begin{adjustbox}{max width=0.5\textwidth}
  \begin{tabular}{ccccccc}
    \toprule
    \multicolumn{5}{r}{Evaluation Metrics}                   \\
    \cmidrule(r){3-7}
    Dataset& Attack Name     & SSIM$\uparrow$  & LPIPS$\downarrow$ &  PSNR$\uparrow$  & MSE$\downarrow$  & RCI$\downarrow$   \\
    \midrule
    &DLG\cite{zhu2019deep} &0.205 &0.439 &9.242 &0.126 &0.441 \\
&Inverting Gradients\cite{geiping2020inverting}  &0.041 &0.587 &5.802 &0.265 & 0.546\\
CIFAR-10 &GradInversion\cite{yin2021see}    &0.645 &0.208 &20.167 &0.009 &0.228 \\
&Multiple Updates\cite{geng2110towards}    &0.395 &0.265 &10.854 &0.089 &0.195 \\

    \midrule
&DLG\cite{zhu2019deep} &0.087 &0.525 &11.379 &0.073 &0.509 \\
&Inverting Gradients\cite{geiping2020inverting} &0.146 &0.660 &8.736 &0.136 &0.619 \\
ImageNet &GradInversion \cite{yin2021see}    &0.246 &0.727 &12.531 &0.061 &0.612 \\
&Multiple Updates \cite{geng2110towards}    & - - - &- - - &- - - &- - - &- - - \\

    \bottomrule
  \end{tabular}
  \end{adjustbox}
\end{table}

\begin{table}
  \caption{The effectiveness of various defence methods  for 10,000 training iterations and using GradInversion every 500 with a batch size of 8 for CIFAR-10 }
  \label{sample-table6}
  \centering
  \begin{adjustbox}{max width=0.5\textwidth}
  \begin{tabular}{ccccccccc}
    \toprule
    \multicolumn{5}{r}{Evaluation Metrics}                   \\
    \cmidrule(r){3-9}
    Dataset& Defence Name        & SSIM$\downarrow$ & LPIPS$\uparrow$  &  PSNR$\downarrow$ & MSE$\uparrow$   & RCI$\uparrow$   & Acc$\uparrow$ & Time(Min)$\downarrow$\\
    \midrule
    &None&0.645  & 0.208 &20.167 &0.009 &0.228 &0.422  &89\\
    \cmidrule(r){2-9}
    &Gaussian Noise(0.1)   &0.015 &0.646&5.538 &0.281 &0.635 &0.391  &99\\
    &Gaussian Noise(0.01)   &0.051 &0.621 &7.523 &0.177 &0.605 &0.407  &104\\
    CIFAR-10&Gaussian Noise(0.001)   & 0.575 &0.270&18.874 &0.014 &0.311 &0.413  &100\\
    &PRECODE\cite{scheliga2022precode}  &0.025 &0.617 &5.813 &0.263 &0.638   &0.389 &52\\
     & DCS\cite{wu2023concealingsensitivesamplesgradient}  &0.073 & 0.561 &11.609 &0.076 &0.516 & 0.362 &233\\

    \bottomrule
  \end{tabular}
  \end{adjustbox}
\end{table}
\begin{figure}
    \begin{minipage}[t]{0.47\textwidth}
    \centering
    \resizebox{0.88\textwidth}{!}{ 
        \begin{tikzpicture}
            \begin{axis}[
                xlabel=$Training Iterations$,
                ylabel=$SSIM$,
                xmin=0, xmax=10000,
                ymin=0, ymax=1,
                xtick={1000,2000,...,10000},
                ytick={0,0.1,...,1},
                legend style={font=\tiny}
            ]
            \addplot[smooth,mark=*,blue] plot coordinates {
(500, 0.23886671)
(1000, 0.239683236)
(1500, 0.230272734)
(2000, 0.219269246)
(2500, 0.213502119)
(3000, 0.228162617)
(3500, 0.233766252)
(4000, 0.242344836)
(4500, 0.236761564)
(5000, 0.235190097)
(5500, 0.240344811)
(6000, 0.222101222)
(6500, 0.249600998)
(7000, 0.258655116)
(7500, 0.256081012)
(8000, 0.237781309)
(8500, 0.254532509)
(9000, 0.261668499)
(9500, 0.261503303)
(10000, 0.259675854)

            }node[above left] {DLG};
            
            \addplot[smooth,color=red,mark=x]
                plot coordinates {
(500, 0.060528968)
(1000, 0.053203556)
(1500, 0.057909214)
(2000, 0.050410093)
(2500, 0.063377686)
(3000, 0.092840024)
(3500, 0.036836412)
(4000, 0.046992937)
(4500, 0.047003426)
(5000, 0.054392943)
(5500, 0.051349804)
(6000, 0.049941384)
(6500, 0.049615498)
(7000, 0.039791591)
(7500, 0.052345589)
(8000, 0.046953112)
(8500, 0.06448878)
(9000, 0.040577251)
(9500, 0.045848312)
(10000, 0.041729043)

                }node[above left] {Inverting Gradients};
            \addplot[smooth,color=green,mark=o]
                plot coordinates {
(500, 0.3711836524307728)
(1000, 0.3981427438557148)
(1500, 0.4341816119849682)
(2000, 0.4877901114523411)
(2500, 0.5401744320988655)
(3000, 0.5234360732138157)
(3500, 0.5678999125957489)
(4000, 0.5649885088205338)
(4500, 0.591597356)
(5000, 0.5661932565271854)
(5500, 0.6186180710792542)
(6000, 0.6102808266878128)
(6500, 0.599786289)
(7000, 0.6450938433408737)
(7500, 0.610294834)
(8000, 0.6227721013128757)
(8500, 0.6327250637114048)
(9000, 0.6561002284288406)
(9500, 0.6293920837342739)
(10000, 0.6452951580286026)

                }node[above left] {GradInversion};
            \addplot[smooth,color=black,mark=triangle]
                plot coordinates {
(500, 0.4902052991092205)
(1000, 0.5211006589233875)
(1500, 0.8521213382482529)
(2000, 0.7001418396830559)
(2500, 0.6642144918441772)
(3000, 0.5247128754854202)
(3500, 0.48117092438042164)
(4000, 0.4898069128394127)
(4500, 0.4380303006619215)
(5000, 0.41927252896130085)
(5500, 0.417415552)
(6000, 0.3987401928752661)
(6500, 0.404635914)
(7000, 0.3932458683848381)
(7500, 0.39964256808161736)
(8000, 0.3896520771086216)
(8500, 0.388446894)
(9000, 0.3890897370874882)
(9500, 0.38320206478238106)
(10000, 0.3849002346396446)
                
                }node[above left] {Multiple Updates};

            \end{axis}
        \end{tikzpicture}
    }
    \caption{SSIM-based reconstruction quality of attack on a CIFAR10 batch of 8 for 4 state-of-the-art attacks across 10,000 training iterations.}
    \label{fig:over times}
  \end{minipage}%
  \hspace{10pt}
\begin{minipage}[t]{0.47\textwidth}
  \centering
  \includegraphics[width=1\linewidth]{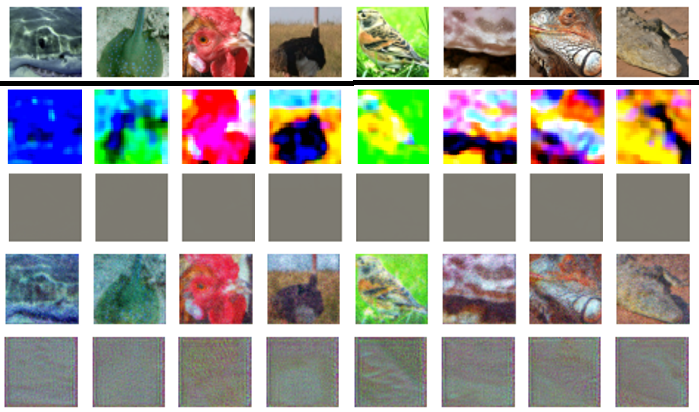}
  \caption{Visual comparison for ImageNet recovery (batch size of 8) with state-of-the-art methods. From top to bottom; Ground Truth, DLG, IG, GradInversion and Multiple Updates.}
  \label{fig:vis}
    \end{minipage}
\end{figure}

\begin{figure}
  \begin{minipage}[t]{0.5\textwidth}
    \centering
    \includegraphics[width=1\linewidth]{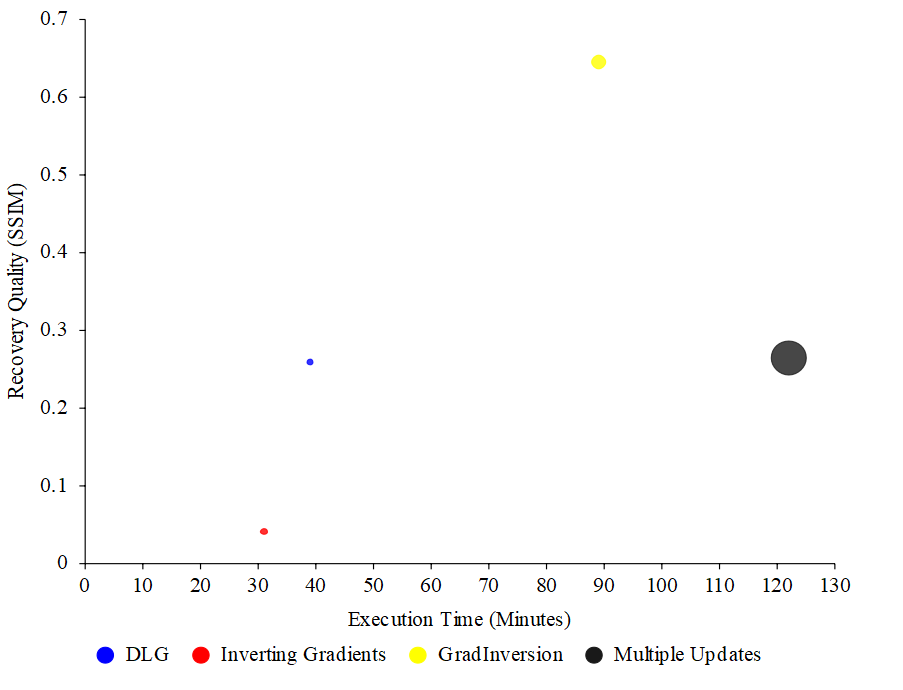}
    \caption{SSIM-based reconstruction quality against execution time with bubble size representing GPU VRAM for 4 state-of-the-art attacks with a batch of 8 CIFAR10 Images.}
    \label{fig:enter-ll}
  \end{minipage}%
  \hspace{10pt}
  \begin{minipage}[t][\height][t]{0.47\textwidth}

    \resizebox{1\textwidth}{!}{ 
            \centering
    \begin{tikzpicture}
        \begin{axis}[
            xlabel={$Batch Size$},
            ylabel={$SSIM$},
            xmin=1, xmax=32,
            ymin=0, ymax=1,
            xtick={1,4,8,16,32},
            ytick={0,0.1,...,1},
            legend style={font=\tiny},
            legend pos=north east,
            grid=both,
            yticklabel style={/pgf/number format/fixed},
        ]
        \addplot[smooth,mark=*,blue] plot coordinates {
            (1,0.323)
            (4,0.240)
            (8,0.213)
            (16,0.119)
            (32,0.064)
        };
        \addlegendentry{DLG}
        
        \addplot[smooth,color=red,mark=x] plot coordinates {
            (1,0.126)
            (4,0.064)
            (8,0.042)
            (16,0.040)
            (32,0.047)
        };
        \addlegendentry{Inverting Gradients}
        
        \addplot[smooth,color=green,mark=o] plot coordinates {
            (1,0.382)
            (4,0.295)
            (8,0.645)
            (16,0.289)
            (32,0.094)
        };
        \addlegendentry{GradInversion}
        
        \addplot[smooth,color=black,mark=triangle] plot coordinates {
            (1,0.993)
            (4,0.442)
            (8,0.383)
            (16,0.385)
            (32,0.382)
        };
        \addlegendentry{Multiple Updates}

        \end{axis}
    \end{tikzpicture}
    }
    \caption{SSIM Comparison for 5 state-of-the-art attacks with different CIFAR10 batch sizes.}
    \label{fig:ssim_batch_comparison}
    \end{minipage}
\end{figure}

\subsection{State-Of-The-Art Attack Comparison}

For each of the 4 attacks across both the CIFAR-10 and ImageNet datasets, we run experiments with an attack batch size of 8 and compare the recovery performances using LPIPS, SSIM, PSNR, MSE, and RCI metrics. Table \ref{sample-table5} presents the results of these experiments. The visualization in Figure \ref{fig:vis} illustrates the recovery performance of ImageNet attacks. DLG and Inverting Gradients tend to perform poorly, with Inverting Gradients appearing excessively faded. The Multiple Updates approach fails on ImageNet as the VRAM required to process dozens of ImageNet weights and gradients exceeded the 80GB limit of the A100. Figure \ref{fig:over times} depicts the SSIM of each attack over 10,000 training iterations. GradInversion’s attack performance typically improves as training progresses because gradient magnitudes decrease. In contrast, Multiple Updates tends to decline in effectiveness over time, as the number of gradient-weight pairs contributing to the loss calculation hits a limit; with each additional pair beyond this threshold, performance diminishes. Inverting Gradients and DLG are generally not very effective, regardless of the training state, and tend to show the best performance in the literature when used to attack gradients from untrained networks.

\subsection{Computational Analysis}

An important aspect when considering the feasibility of a gradient inversion attack is the computation cost. However, no previous work has provided these figures, making it challenging to contrast different approaches. Figure \ref{fig:enter-ll} illustrates the evaluation of all 4 attacks on a batch of 8 CIFAR10 images, examining GPU VRAM usage, execution time, and SSIM recovery performance. DLG and Inverting gradients are significantly more efficient than the other attacks, but the recovery quality is much lower than GradInversion. The computational demands of GradInversion scales with the number of optimized random seeds and Multiple Updates scales with the number of steps employed. As outlined above, the computation cost of Multiple Updates makes the attack infeasible when using more complex datasets, hence why these results were unobtainable in Table \ref{sample-table5}.

\subsection{Effect of Increasing Batch Size}

To accommodate varying batch sizes, the regularization scaling factors are dynamically adjusted, as outlined in Table \ref{sample-table4}. Our experiments examined CIFAR-10 batches up to size 32 for each attack, as illustrated in Figure \ref{fig:ssim_batch_comparison}. Generally, SSIM recovery decreases as batch size increases across all four attacks. Notably, GradInversion diverges from this trend by performing better at a batch size of 8 before dropping to levels similar to Inverting Gradients and DLG. We attribute this to GradInversion’s improved performance with increased training (see Figure \ref{fig:over times}), where a batch size of 8 reaches lower-magnitude gradients faster. However, this effect doesn’t extend to batch sizes of 16 or 32, as GradInversion’s label recovery is most effective with batches containing unique labels, which is less likely at larger batch sizes in CIFAR-10. In contrast, Multiple Updates achieve strong performance for smaller batch sizes due to closely aligned weight pairs in the loss calculation. When weight pairs span a broader range of training, Multiple Updates’ performance declines.

\subsection{State-Of-The-Art Defence Comparison}

Table \ref{sample-table6} evaluates various defense methods on CIFAR-10 over 10,000 training iterations, using GradInversion every 500 iterations with a batch size of 8. Increasing Gaussian noise enhances privacy but leads to a significant decline in classification accuracy. The highest intensity of Gaussian noise results in the lowest accuracy, indicating a trade-off between privacy and performance. In contrast, PRECODE maintains similar privacy and model accuracy while reducing computation. This is because the attack converges on a suboptimal solution much earlier. DCS considers an optimization defence strategy and has relatively poor privacy, model accuracy and execution time. Overall, these results highlight the importance of considering accuracy alongside privacy when evaluating defense strategies.
\section{Conclusions}
 This paper introduces a comprehensive framework called FEDLAD for evaluating Deep Leakage attacks and defenses within the context of Federated Learning, advancing the understanding of security challenges associated with this decentralized paradigm. By implementing a unified benchmark that encompasses multiple state-of-the-art gradient inversion techniques alongside various defense strategies, we provide a robust platform for researchers to assess and compare the efficacy of these methods across different datasets and training states. Our findings indicate that GradInversion performs best under specific conditions, while DLG and Inverting Gradients show lower effectiveness. The Multiple Updates method struggles with high computational demands, limiting its feasibility on complex datasets. Furthermore, we highlight a crucial trade-off between privacy and model accuracy, particularly with Gaussian noise. Overall, this work aims to stimulate future research, enhance reproducibility, and contribute to the development of more secure and efficient decentralized machine learning systems.
{
    \small
    \bibliographystyle{ieeenat_fullname}
    \bibliography{main}
}


\end{document}